\documentclass[twocolumn,showpacs,preprintnumbers,amsmath,amssymb]{revtex4}
\usepackage{graphicx}% Include figure files
\usepackage{dcolumn}% Align table columns on decimal point
\usepackage{bm}% bold math

\begin{document}

\title{Macroscopic quantum coherence in spinor condensates confined in an anisotropic potential}
\author{Yixiao Huang$^{1,4}$, Yunbo Zhang$^{2}$, Rong L\"{u}$^{3}$, Xiaoguang Wang$^{1}$, and Su Yi$^{4}$}
\affiliation{$^{1}$Zhejiang Institute of Modern Physics, Department of Physics, Zhejiang University, Hangzhou 310027, China}
\affiliation{$^{2}$2Department of Physics and Institute of Theoretical Physics, Shanxi University, Taiyuan 030006, China}
\affiliation{$^{3}$Department of Physics, Tsinghua University, Beijing 100084, China}
\affiliation{$^{4}$State Key Laboratory of Theoretical Physics, Institute of Theoretical Physics, Chinese Academy of Sciences, P.O. Box 2735, Beijing 100190, China}
\date{\today}


\begin{abstract}
We investigate the macroscopic quantum coherence of a spin-1 Rb condensate confined in an anisotropic potential. Under the single-mode approximation, we show that the system can be modeled as a biaxial quantum magnet due to the interplay between the magnetic dipole-dipole interaction and the anisotropic potential. By applying a magnetic field along the hard-axis, we show that the tunneling splitting oscillates as a function of the field strength. We also propose an experimental scheme to detect the oscillatory behavior of the tunneling splitting by employing the Landau-Zener tunneling.
\end{abstract}
\volumeyear{year}
\volumenumber{number}
\issuenumber{number}
\eid{identifier}
\startpage{1}
\endpage{2}
\maketitle

\section{Introduction}
Tunneling of a macroscopic variable into a classically forbidden region provides one of the most striking manifestations of quantum mechanics~\cite{legg}. The model of quantum tunneling often involves a particle moving in a multistable potential. Quantum mechanically, there is a finite probability for the particle to tunnel through the barrier and escape from a metastable state to an absolutely stable one, which is often referred to as the macroscopic quantum tunneling (MQT). In a symmetric double-well, the particle tunnels through the barrier to oscillate back and forth between the degenerate states, known as macroscopic quantum coherence (MQC). The quantum tunneling removes the degeneracy of the ground states, resulting a tunneling splitting between the true ground state and the first excited state. 

Because of its small size and precise characterizability, single-molecule nanomagnet represents an ideal platform for demonstrating the MQT and MQC of the spin~\cite{gunther,chudn,miller,fried}. For such a system, MQT consists of the tunneling of the magnetization out of the metastable easy directions in the presence of external field, which was experimentally observed as a series of steps in the hysteresis loops in Mn$_{12}$-acetate and Fe$_{8}$ molecules at low temperatures~\cite{fried2,hern,thomas,sang}. MQC in spin systems represents the resonance between two equivalent easy directions. Of particular interest, in the present of a general biaxial anisotropy, the spin has two preferred tunneling path via the medium axis when the external field is applied along the hard direction. As a result, the constructive and destructive interference then give rise to the oscillatory tunneling splitting~\cite{garg}. The geometric-phase effect of the quantum spin tunneling was first observed by Wernsdorfer and Sessoli~\cite{werns}. 

In the context of the ultracold atomic gases, the MQC problem was previously studied for a spinor condensate trapped in a double-well potential, where the magnetic dipole-dipole interaction (MDDI) between the condensates confined in different wells induces an uniaxial anisotropy with the easy-axis being along the direction connecting two potential wells~\cite{pu}. In fact, even in a single axially symmetric trap, a dipolar spinor condensate can be treated as an uniaxial quantum magnet~\cite{yi}, whose magnetic properties, spin squeezing, magnetization steps, and macroscopic entanglement generation, and MQC were studied~\cite{yi1,zhang}. However, for uniaxial quantum magnet, the tunneling splitting of model is a monotonically increasing function of the transverse field strength~\cite{pu}. The possibility of obtaining the oscillatory tunneling splitting in ultracold atomic gases was explored by considering a condensate coupled dispersively with an ultrahigh-finesse optical cavity~\cite{cheng}.

In the present work, we investigate the MQC of a spin-1 Rb condensate confined in a three-dimensional anisotropic harmonic oscillator potential. Under the SMA, we showed that the interplay of the MDDI and the anistropic trap results in a biaxial quantum magnet whose magnetic structure can be approximately specified by the geometry of the trapping potential. Subsequently, we study the MQC of of the condensate by applying an external magnetic field along the hard-axis. Similar to the molecular magnet, we show that the tunneling splitting of our system oscillates as a function of the field strength.  Finally, we propose an experimental scheme to detect the MQC by utilizing the Landau-Zener transition.

This paper is organized as follows. In Sec.~\ref{secmod}, under the single-mode approximation (SMA), we derive the Hamiltonian of the system. In Sec.~\ref{secmags}, we explore the magnetic structure of the system. Section~\ref{secmqc} is devoted to the properties of the MQC and its experimental detection. Finally, we conclude in Sec.~\ref{secconcl}

\section{Model}\label{secmod}
We consider a trapped gas of $N$ spin $F=1$ Rb atoms subjected to an external magnetic field ${\mathbf B}$. Atoms interact via the $s$-wave collisions and the MDDI. In the second quantized form, the total Hamiltonian of the system reads~\cite{yi}
\begin{widetext}
\begin{eqnarray}
\mathcal{H}&=&\int d\mathbf{r}\hat{\psi}_{\alpha}^{\dagger}
(\mathbf{r})\left[\left(-\frac{\hbar^{2}\nabla^{2}}{2M}+V_{\text{ext}
}(\mathbf{r})\right)\delta_{{\alpha\beta}}-g_{F}\mu_{B}{\mathbf B}\cdot{\mathbf F}_{\alpha\beta}\right]\hat{\psi}_{\beta}(\mathbf{r})\nonumber\\
&&+\frac{c_{0}}{2}\int d\mathbf{r}\hat{\psi}_{\alpha}^{\dagger
}(\mathbf{r})\hat{\psi}_{\beta}^{\dagger}(\mathbf{r})\hat{\psi}_{\alpha
}(\mathbf{r})\hat{\psi}_{\beta}(\mathbf{r})+\frac{c_{2}}{2}\int d\mathbf{r}\hat{\psi}_{\alpha}^{\dagger}
(\mathbf{r})\hat{\psi}_{\alpha^{\prime}}^{\dagger}(\mathbf{r})\mathbf{F}
_{\alpha\beta}\cdot\mathbf{F}_{\alpha^{\prime}\beta^{\prime}}\hat{\psi}
_{\beta}(\mathbf{r})\hat{\psi}_{\beta^{\prime}}(\mathbf{r})\nonumber\\
&&+\frac{c_{d}}{2}\int \frac{d\mathbf{r} d\mathbf{r}^{\prime
}}{\left\vert \mathbf{r}-\mathbf{r}^{\prime}\right\vert ^{3}
}{\bigg [}\hat{\psi}_{\alpha}^{\dagger}(\mathbf{r})\hat{\psi}
_{\alpha^{\prime}}^{\dagger}(\mathbf{r}^{\prime})\mathbf{F}_{\alpha\beta}
\cdot\mathbf{F}_{\alpha^{\prime}\beta^{\prime}}\hat{\psi}_{\beta}
(\mathbf{r})\hat{\psi}_{\beta^{\prime}}(\mathbf{r}^{\prime})-3\hat{\psi}_{\alpha}^{\dagger}(\mathbf{r})
\hat{\psi}_{\alpha^{\prime}}^{\dagger}(\mathbf{r}^{\prime
})(\mathbf{F}_{\alpha\beta}\cdot\mathbf{e)(F}_{\alpha^{\prime}\beta^{\prime}
}\cdot\mathbf{e)}\hat{\psi}_{\beta}(\mathbf{r})\hat{\psi}_{\beta^{\prime}
}(\mathbf{r}^{\prime}){\bigg ]},
\label{hami}
\end{eqnarray}
\end{widetext}
where $\hat{\psi}_{\alpha}$ is the field operator for $m_{F}=\alpha$ spin state, $V_{\rm ext}$ is the anisotropic confining potential, ${\mathbf F}$ is the angular momentum operator, $g_{F}$ is the Land\'{e} $g$-factor, and $\mu_{B}$ is the Bohr magneton. The spin-independent and spin-exchange $s$-wave collsions are characterized by $c_{0}=4\pi\hbar^{2}(a_{0}+2a_{2})/(3M)$ and $c_{2}=4\pi\hbar^{2}(a_{2}-a_{0})/(3M)$, respectively, with $a_{f=0,2}$ being the scattering length of two spin-1 atoms in the combined symmetric channel of the total spin $f$~\cite{ho,machida}. In particular, we have $c_{2}<0$ for Rb atoms, indicating that the spin-exchange interaction is ferromagnetic. The strength of the MDDI is $c_{d}=\mu_{0}\mu_{B}^{2}g_{F}^{2}/(4\pi)$ with $\mu_{0}$ being the vacuum permeability.

To proceed further, we adopt the SMA which assumes that atoms in different spin states share a common spatial mode function $\phi(\mathbf{r})$. The field operators can then be decomposed into~\cite{law}
\begin{equation}
\hat{\psi}_{\alpha}(\mathbf{r})\simeq\phi(\mathbf{r})\hat{a}_{\alpha},\label{sma}
\end{equation}
where $\hat{a}_{\alpha}$ being the annihilation operator of the spin component $\alpha$. The validity of the SMA can be justified by noting that $c_{0}\gg |c_{2}|$ and $c_{d}\simeq 0.1|c_{2}|$ for Rb atom~\cite{yi2}. By substituting Eq. (\ref{sma}) into (\ref{hami}) and dropping the constant terms, the total Hamiltonian reduces to
\begin{align}
{\cal H}=&\,(c_{2}'-c_{d}')\hat{\mathbf S}^{2}+3c_{d}'\hat S_{z}^{2}-3c_{d}''(\hat S_{x}^{2}-\hat S_{y}^{2})-g_{F}\mu_{B}{\mathbf B}\cdot\hat{\mathbf S}\nonumber\\
&+3c_{d}^{\prime}\hat{a}_{0}^{\dagger}\hat{a}_{0}+3c_{d}^{\prime\prime}(\hat{a}_{-1}^{\dagger}\hat{a}_{1}+\hat{a}_{1}^{\dagger}\hat{a}_{-1}), \label{hsma}
\end{align}
where $\hat{\mathbf S}=\sum_{\alpha\beta}\hat a_{\alpha}^{\dag}{\mathbf F}_{\alpha\beta}a_{\beta}$ is the total many-body angular momentum operator and $\hat S_{\eta}$ ($\eta=x,y,z$) is its projection along $\eta$-axis, $c_{2}'=(c_{2}/2)\int d{\mathbf r}|\phi({\mathbf r})|^{4}$ is the strength of spin-exchange interaction, and the strength of the MDDI are characterized by two parameters:
$c_{d}'=(c_{d}/4)\int d\mathbf{r}d\mathbf{r}^{\prime}|\phi(\mathbf{r}
)|^{2}|\phi(\mathbf{r}^{\prime})|^{2}\left\vert \mathbf{r}-\mathbf{r}
^{\prime}\right\vert ^{-3}\left(1-3\cos^{2}\vartheta\right)$ and 
$c_{d}''=(c_{d}/4)\int d\mathbf{r}d\mathbf{r}^{\prime}|\phi(\mathbf{r}
)|^{2}|\phi(\mathbf{r}^{\prime})|^{2}\left\vert \mathbf{r}-\mathbf{r}
^{\prime}\right\vert ^{-3}\sin^{2}\vartheta \,e^{2i\varphi}$
with $\vartheta$ and $\varphi$ being the polar and azimuthal angles of the vector ${\mathbf r}-{\mathbf r}'$, respectively. It can be shown that $c_{d}''=0$ if the mode function $\phi({\mathbf r})$ possesses an axial symmetry~\cite{yi}. One should note that the last line of the Eq. (\ref{hsma}) originates from the commutation relations between the bosonic operators.

\begin{figure}[ptb]
\includegraphics[width=3.2in]{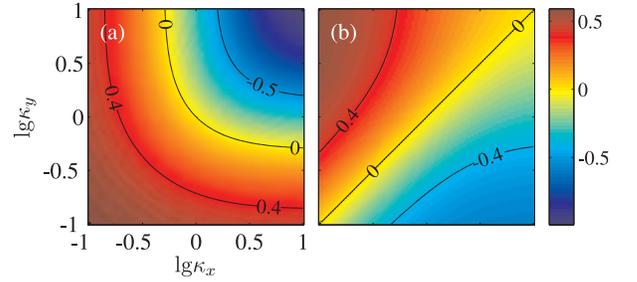}
\caption{(Color online). Anisotropic constants $D(\kappa_{x},\kappa_{y})$ (a) and $E(\kappa_{x},\kappa_{y})$ (b) for Rb condensates.}
\label{coeffde}
\end{figure}

It is convenient to rescale the Eq. (\ref{hsma}) by using $|c_{2}^{\prime}|$ as the energy unit, which yields the dimensionless Hamiltonian
\begin{align}
\mathcal{H}&={\cal H}_{0}+{\cal H}',\label{htot}\\
{\cal H}_{0}&=-\frac{3+D}{3}\hat{\mathbf S}^{2}-D\hat S_{z}^{2}+E(\hat S_{x}^{2}-\hat S_{y}^{2})-{\mathbf H}\cdot\hat {\mathbf S},\label{hzero}\\
{\cal H}'&=-D\hat{a}_{0}^{\dag}\hat a_{0}-E\left(\hat{a}_{-1}^{\dagger}\hat{a}_{1}+\hat{a}_{1}^{\dagger}\hat{a}_{-1}\right),\label{hone}
\end{align}
where $D=-3c_{d}'/|c_{2}^{\prime}|$ and $E=-3c_{d}''/|c_{2}^{\prime}|$ are the anisotropy constants and ${\mathbf H}=g_{F}\mu_{B}{\mathbf B}/|c_{2}'|$ is the dimensionless magnetic field. Apparently, the Hamiltonian (\ref{hzero}) describes a biaxial quantum magnet. To be more specific, we assume that the mode function is a Gaussian
$\phi(\mathbf{r})=\pi^{-3/4}(q_{x}q_{y}q_{z})^{-1/2}e^{-\sum_{\eta=x,y,z}\eta^{2}/(2q_{\eta}^{2})}$ 
with $q_{\eta}$ being the width of the condensate along the $\eta$ direction. It can be shown that 
\begin{align}
D(\kappa_{x},\kappa_{y})=&-\frac{4\pi c_{d}}{|c_{2}|}\kappa_{x}\kappa_{y}\int dt\,t
e^{-(\kappa_x^2+\kappa_y^2)t^2/2}\nonumber\\
&\times I_0\left(\frac{1}{2}(\kappa_x^2-\kappa_y^2)t^2\right)
\left[2-3\sqrt{\pi}\,te^{t^2}{\rm
erfc}(t)\right],\nonumber\\
E(\kappa_{x},\kappa_{y})=&-\frac{4\pi^{3/2}c_{d}}{|c_{2}|}\kappa_{x}\kappa_{y}\int
dt\,t^2e^{-(\kappa_x^2+\kappa_y^2)t^2/2}\nonumber\\
&\times I_1\left(\frac{1}{2}(\kappa_x^2-\kappa_y^2)t^2\right)e^{t^2}{\rm
erfc}(t),\nonumber
\end{align}
where the set of parameters $(\kappa_{x},\kappa_{y})\equiv (q_{x}/q_{z},q_{y}/q_{z})$ characterizes the geometry of the condensate, $I_{0,1}(\cdot)$ is the modified Bessel functions of the first kind, and ${\rm erfc}(\cdot)$ is the complementary error function. Figure~\ref{coeffde} shows the values of $D(\kappa_{x},\kappa_{y})$ and $E(\kappa_{x},\kappa_{y})$ for Rb condensate. In particular, $E(\kappa_{x},\kappa_{y})=0$ when $\kappa_{x}=\kappa_{y}$, the condensate becomes an uniaxial magnet~\cite{yi,yi1}.

\begin{figure}[ptb]
\includegraphics[width=2.2in]{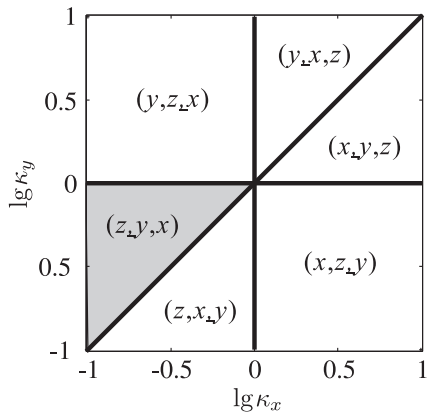}
\caption{Magnetic structure of a Rb condensate on the parameter $(\kappa_{x},\kappa_{y})$ plane in terms of (easy, medium, hard) axes in the absence of magnetic field. The solid lines correspond to uniaxial magnets.}
\label{mphase}
\end{figure}

\section{Magnetic structure}\label{secmags}
To find the ground state wave function, we have numerically diagonalized the Hamiltonian (\ref{htot}) in the Fock state basis, $\{|N_{1},N_{0},N_{-1}\rangle\}$ subjected to the constrains $N_{\alpha}\geq0$ and $\sum_{\alpha}N_{\alpha}=N$, for the parameter regime covering $N\leq 50$, $0.1\leq \kappa_{x,y}\leq 10$, and arbitrary magnetic field. It is found that, as long as $N\geq 5$, we always have $\langle {\mathbf S}^{2}\rangle\simeq N(N+1)$, indicating that the condensate stays in a ferromagnetic state with total spin $S\simeq N$. In addition, the contribution from ${\cal H}'$ [Eq.~(\ref{hone})] is negligible. Consequently, we may drop the constant ${\mathbf S}^{2}$ term in ${\cal H}_{0}$ and the linear term ${\cal H}'$ such that the Eq.~(\ref{htot}) reduces to the familiar biaxial Hamiltonian
\begin{eqnarray}
{\cal H}_{\rm eff}\simeq -D\hat S_{z}^{2}+E(\hat S_{x}^{2}-\hat S_{y}^{2})-{\mathbf H}\cdot\hat {\mathbf S}\label{hreduce}
\end{eqnarray}
for quantum magnet with total spin $S=N$. 

In the absence of the external magnetic field, the magnetic properties of the reduced Hamiltonian (\ref{hreduce}) are completely determined by the anisotropic constants $D$ and $E$. In Fig.~\ref{mphase}, we classify the magnetic structure of the system on the geometric parameter plane in terms of the easy, medium, and hard axes, which essentially states that the easy (hard) axis corresponds to the direction with the weakest (strongest) confinement. Figure~\ref{mphase} can be physically understood as follows. Without the MDDI, the ground state energy is degenerate about the orientation of the spin. The MDDI removes this degeneracy by pointing the spins to certain directions. Noting that, for two magnetic dipoles, the dipolar interaction is attractive (repulsive) for the head-to-tail (side-by-side) configuration. Therefore, the spins prefer to align in the way such that the possibility for the head-to-tail configuration is maximized, which is exactly the direction corresponding to the weakest confinement.

\section{Macroscopic quantum coherence}\label{secmqc}
A simple model for studying the MQC problem is an uniaxial magnet with easy-axis anisotropy, which can be realized, for example, by taking $D<0$ and $E=0$ in the Hamiltonian (\ref{htot}). For ultracold atomic gases, the MQC of such model was previously studied in the Refs.~\cite{pu,zhang}. Here, without loss of generality, we study the MQC of a biaxial magnet with $D>E>0$, which corresponds to the geometric parameters $(\kappa_{x},\kappa_{y})$ in the shaded region of Fig.~\ref{mphase}. The magnetic field is applied along the hard axis, i.e., ${\mathbf H}=H_x \hat {\mathbf x}$.

We first consider the classical counterpart of the reduced Hamiltonian ({\ref{hreduce}) by treating $\hat{\mathbf S}$ as a vector of length $|{\mathbf S}|=N$. The total energy then becomes
\begin{align}
{\cal E}(\theta,\phi)=&-DN^{2}\cos^{2}\theta+EN^{2}\sin^{2}\theta\cos2\phi\nonumber\\
&-H_{x}N\sin\theta\cos\phi,\nonumber
\end{align}
where $(\theta,\phi)$ denotes the direction of ${\mathbf S}$. Apparently, for $H_{x}<H_{x}^{*}\equiv 2N(D+E)$, there exist two degenerate ground states located at $(\theta_{0},0)$ and $(\pi-\theta_{0},0)$ with $\sin\theta_{0}=H_{x}/[2N(D+E)]$. When $H_{x}\geq H_{x}^{*}$, the double degeneracy is removed such that the system is fully polarized along the $x$ axis by the external field. 

\begin{figure}[ptb]
\includegraphics[width=2.8in]{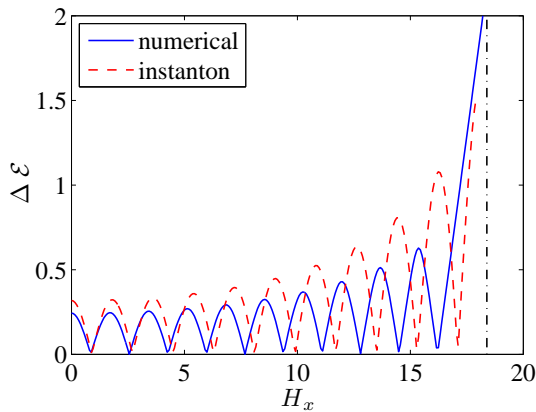}
\caption{(Color online). Tunneling splitting as a function of external magnetic field for $\lg\kappa_{x}=-0.9$, $\lg\kappa_{y}=-0.1$, and $N=10$. The vertical dash-dotted line marks the position of the critical field.}
\label{dele}
\end{figure}

Quantum mechanically, the classical degeneracy of the ground states is lifted by quantum tunneling even when $H_{x}<H_{x}^{*}$, which results a tunneling splitting $\Delta{\cal E}$ between the true ground state and the first excited state. Of particular interest, with the biaxial anisotropy, there exist two tunneling paths with opposite windings on the $yz$ easy anisotropy plane. The constructive and destructive interferences of quantum spin phases of the different paths cause the tunneling splitting to oscillate with the magnetic field. This phenomenon was first predicted by Garg~\cite{garg} and was experimentally observed by Wernsdorfer and Sessoli in Mn$_{12}$ molecules~\cite{werns}. Using the instanton method, the tunneling splitting can be expressed analytically as~\cite{garg,kou}
\begin{equation}
\Delta {\cal E}=\Delta\varepsilon_{0}|\cos(\pi\Theta)|,\label{dele2}
\end{equation}
where $\Theta(H_{x})=N-H_{x}/\left[2\sqrt{2E(D+E)}\right]$ is the area on the Bloch sphere enclosed by the two instanton paths and $\Delta\varepsilon_{0}$ is the tunneling splitting under zero external field which contains the contributions from the classical action and the fluctuations around the instanton paths~\cite{garg,kou}. Clearly, the quantum tunneling is completely quenched whenever $\Theta=n+1/2$ with $n$ being an integer. The period of this oscillation is~\cite{garg}
\begin{eqnarray}
\Delta H_{x}=2\sqrt{2E(D+E)}.
\end{eqnarray}
We want to emphasize that the instanton method is only valid for $H_x<H_x^*$ as it depends on the classical paths of the tunneling.

In Fig.~\ref{dele}, we present the field dependence of the tunneling splitting obtained via the exact numerical diagonalization of the full Hamiltonian (\ref{htot}) for $N=10$, $\lg\kappa_x=-0.9$, and $\lg\kappa_y=-0.1$. As a comparison, we also plot Eq. (\ref{dele2}) with $\Delta\varepsilon_{0}$ from Ref.~\cite{kou}. As can be seen, for small $H_x$, the agreement between two approaches is quite well, which further confirms that our system are well described by the reduced Hamiltonian. However, significant discrepancy is observed for the larger external field. It is found that, by increasing the atom number $N$, the agreement between the numerical and instanton methods can be improved. 

\begin{figure}[ptb]
\includegraphics[width=2.6in]{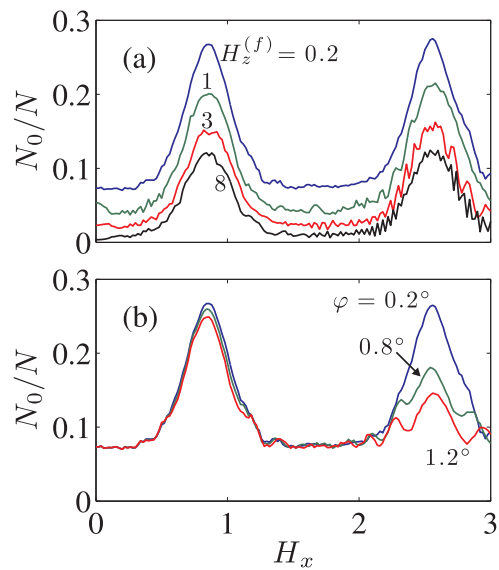}
\caption{(Color online). $N_{0}/N$ as a function of $H_{x}$ for $\varphi=0$ (a) and $\varphi\neq 0$ with $H_{z}^{(f)}=0.2$ (b). Other parameters are $N=10$, $\lg\kappa_{x}=-0.9$, $\lg\kappa_{y}=-0.1$, $v=10^{-3}$, and $H_{z}^{(i)}=-8$.}
\label{mn0}
\end{figure}

We now turn to discuss the experimental detection of the MQC in spin-1 Rb condensates. Here we adopt a similar scheme to that used in the molecular magnet experiment by utilizing the Landau-Zener transition~\cite{werns}. More specifically, the experiment can be carried out as follows. In addition to the constant transverse field $H_{x}\hat{\mathbf x}$, we introduce a time-dependent longitudinal magnetic field $H_{z}(t)\hat{\mathbf z}$ which is swept linearly with a constant rate $v>0$, i.e., $H_z(t)=H_z^{(i)}+vt$ for $t\geq0$. When the longitudinal field reaches designated $H_{z}^{(f)}\equiv H_{z}^{(i)}+vt_{f}>0$ at $t=t_{f}$, one measures the number of atoms $N_{\alpha}$ in each spin component. 

As can be seen from the reduced Hamiltonian (\ref{hreduce}), if the initial longitudinal field $H_{z}^{(i)}$ ($<0$) is sufficiently large, the initial state roughly stays at the $S_{z}=-N$ level which anticrosses with the $S_{z}=N$ level at $H_{z}=0$. When the longitudinal field sweeps through this avoid crossing, a Landau-Zener transition occurs. Since the energy gap between the levels $S_{z}=-N$ and $N$ at the anticrossing is exactly the tunneling splitting $\Delta{\cal E}$ which depends on the transverse field, it is expected that $N_{\alpha}$ will oscillate as a function of $H_{x}$. 

The proposed experiment can be simulated by numerically evolving the full Hamiltonian (\ref{htot}) with an initial state being the ground state under the initial magnetic field. Figure~\ref{mn0}(a) shows the typical transverse field dependence of $N_{0}$ for various final longitudinal fields. As expected, $N_{0}$ oscillates as a function of the transverse field strength. The oscillation period is also in very good agreement with that of $\Delta{\cal E}$. In addition, by increasing $H_{z}^{(f)}$, more anticrossings in the energy spectrum of the Hamiltonian (\ref{htot}) are swept through. As a result, the oscillation amplitude decreases with $H_{z}^{(f)}$, and eventually the $N_{0}(H_{x})$ will roughly converge to the curve corresponding to $H_{z}^{(f)}=8$ as one further increases $H_{z}^{(f)}$. Experimentally, it is possible that the transverse field is misaligned such that it forms an angle $\varphi$ to the $x$-axis. In Fig.~\ref{mn0}(b), we plot $N_{0}(H_{z})$ for $\varphi\neq 0$. The oscillation remains for small $\varphi$, however, it will disappear for large $\varphi$. Finally, we remark that $N_{1}$ and $N_{-1}$ also exhibit the similar oscillatory behavior as $N_{0}(H_{z})$. 

\section{Conclusion}\label{secconcl}
To conclude, we have studied the MQC of a spin-1 Rb condensate confined in an anisotropic trap. Under the SMA, we showed that this system can be described as a biaxial quantum magnet. Physically, this biaxial anisotropy is induced by the MDDI and the anisotropic trap. Subsequently, we showed that the magnetic structure of the system is determined by the geometric parameters of the trapping potential. We then studied the MQC of the spinor condensate by applying an external magnetic along the hard-axis, we showed that the tunneling splitting oscillates as a function of the field strength. Finally, we propose to detect the MQC by utilizing the Landau-Zener transition.

\section*{ACKNOWLEDGMENTS}
SY acknowledges support by the NSFC (Grant No. 11025421) and the National 973 program (Grant No. 2012CB922104). XW acknowledges support by the NSFC (Grant No. 11025527) and NFRPC (Grant No. 2012CB921602). YH acknowledges support by the NSFC (Grant No. 10935010). YZ acknowledges support by the NSFC (Grants No. 11074153).

\end{document}